\documentclass[%
 aip,
 cha,%
 amsmath,amssymb,
 reprint,%
]{revtex4-1}
\usepackage{color}
\usepackage{graphicx}
\usepackage{dcolumn}
\usepackage{bm}
\usepackage{amsfonts}
\usepackage{sidecap}
\usepackage{float}
\usepackage{parskip}
\usepackage{grffile}

\usepackage{color}
\usepackage{hyperref}
\usepackage{hhline}
\usepackage{mathtools}
\usepackage[normalem]{ulem}
\makeatletter
\def\@eqnnum{{\normalsize \normalcolor (\theequation)}}
 \makeatother
\begin{document}

\title{Is repulsion good for the health of Chimeras?}
\author{Sarika Jalan}
\email{sarikajalan9@gmail.com}
\author{Saptarshi Ghosh}
\email{sapta15@gmail.com}
\author{Bibhabasu Patra}
\affiliation{Complex Systems Lab, Discipline of Physics, Indian Institute of Technology Indore, Simrol, Indore 453552}
\date{\today}

\begin{abstract}
Yes! Very much so.\\

\noindent{A chimera state refers to the coexistence of a coherent-incoherent dynamical evolution of identically
coupled oscillators. We investigate the impact of multiplexing of a lyer having repulsively coupled oscillators on
occurrence of chimeras in the layer having attractively coupled identical oscillators.
We report that  there exists an enhancement in the appearance of chimera state in one layer of multiplex network in the presence of repulsive coupling in the other layer. Furthermore, we show that a small amount of inhibition or repulsive coupling in one layer is sufficient to yield chimera state in another layer by {destroying {its} synchronized behavior}. These results can be used to get insight into dynamical behaviors of those systems where both
attractive and repulsive coupling exist among their constituents.}

\end{abstract}
\maketitle

\begin{quotation}
In 1975, Kuramoto had introduced a mathematical model for nonidentical, nonlinear phase oscillators which {exhibits a convergence} to a global synchrony at a critical coupling value. In 2002, Kuramoto demonstrated that identical phase oscillators can {show coexistence of} coherent and incoherent dynamics under certain special conditions. Later, Abrams and Strogatz, christened {this} dynamical state as chimera state and provided a detailed description about special conditions required for {its emergence}. Since then, chimera remains an exotic phenomenon with ambiguities. Initially, non-local, non-global topology was stated to be a prerequisite for the emergence of the chimera state. {However, subsequent} investigations {presented an} appearance of the chimera state for both local and global couplings.

{Furthermore, multiplex framework which incorporate various types of interactions between the same pair of nodes as different layers, provide a better portrayal of complex natural networks.} Here we consider a multiplex network with each layer having identical coupling architecture but not necessarily {identical nature of the couplings. Particularly, we consider attractive and repulsive coupling among dynamical units and show} that by keeping network architecture same, coupling in one layer {have profound impact on} the occurrence of chimera in another layer in multiplex networks. This investigation is crucial for many different {complex systems possessing different types of coupling, particularly attractive and repulsive coupling among the same units}. For instance, the brain has inhibitory and excitatory neurons representing {repulsive and attractive} couplings, respectively. Studies on the impact of inhibition on the emergence of chimera state in the multiplex framework will be useful for a better understanding of such complex systems in different conditions.
\end{quotation}

\section{Introduction}
In past few decades study of networks has been catapulted into the limelight due to their booming applications in various branches of science and engineering~\cite{network_rev}. One of the aspects of network science is to study the collective output of a system arising due to complex interactions between its constituents~\cite{sync_rev}. The interactions (edges) between the constituents (nodes) often result in fascinating spatiotemporal phenomena among which hybrid patterns of chimera deserves particular attention. Similar to its namesake in Greek mythology, a chimera state consists of two or more spatial coherent and incoherent domains in a network with symmetric coupling. Chimera state had been initially reported for a ring of identical phase oscillators with non-local coupling~\cite{chim_def}. Since then, a vast amount of theoretical and experimental investigations on chimeras have been performed in a diverse range of oscillators~\cite{chim_rev1}. The investigations resulted in plethora of different variants of chimera including multi-cluster chimera~\cite{mult_clus}, virtual~\cite{vir_chim}, breathing~\cite{breathing_c}, traveling~\cite{travelling_c}, globally clustered chimera~\cite{gcc_chim} and two-dimensional chimera~\cite{2D}. Recently, chimera state is shown for both global~\cite{chim.global} and local coupling~\cite{chim.local_coup}, hyperbolic chaotic oscillators~\cite{hyperbol_chim} as well as for random initial conditions~\cite{Random_init_chim}, negating previous constraints on the emergence of chimera state.

\begin{figure}[t]
 \centerline{\includegraphics[width=3.2in,height=1.2in]{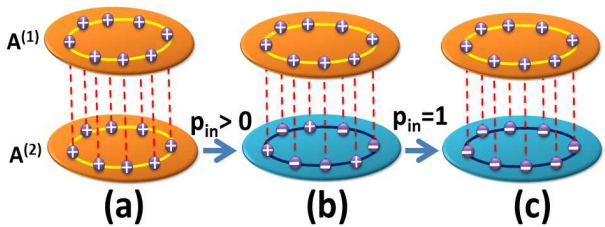}}
 \caption{(Color online) Schematic diagram {of} multiplex network consisting of two layers. Each layer is represented {by} identical 1D lattice  ($S^1$ ring), where each node ({open} circle) has the same coupling architecture. In the second layer of the multiplex network, nodes are (a) attractively coupled, (b) repulsively coupled with probability $p_{in}$, (c) all repulsively coupled corresponding to $p_{in}=1$. Inter layer connections {between two layers are} represented as dashed lines.}
 \label{multiplex}
 \end{figure}

Furthermore, multiplex network framework incorporates existence of
various types of interactions between the same pair of nodes by categorizing them in different layers
with each layer reflecting a specific type of interaction \cite{mul_def}. This framework provides a better
representation of many real-world systems~\cite{mul_real}. Influence of multiple types of interaction on dynamical behavior of an individual layer has been one of the prime objectives of the studies
on multiplex networks~\cite{multi_appl}. The chimera state in multiplex networks
has been reported to occur first time in Ref.~\cite{chim.multiplex1} followed by other investigations \cite{chim.multiplex2,chim_DGhosh}. In this article, we sought to examine the impact of repulsive couplings on chimera patterns in multiplex networks (Fig.~\ref{multiplex}).

For sufficiently high couplings, an array of coupled phase oscillators has been shown to converge to a mean frequency exhibiting exact synchrony under attractive coupling~\cite{k_osc}. Whereas, repulsive couplings have been shown to affect dynamical evolution of nonlinear units in different ways.
The repulsive couplings may either destroy a synchronized state leading to the
incoherent evolution~\cite{repsl_k_osc} or lead to an
emergence of new synchronization regimes in addition to the multistability phenomenon \cite{repulsive_new}. An emergence of the chimera state under attractive and repulsive couplings in a globally coupled monoplex network  has already been reported \cite{ar_chim}. Here we investigate impact of repulsive couplings on the occurrence of
chimera state in multiplex networks. Specifically, we demonstrate that the emergence and presence of
chimera state in one layer can be controlled by changing the probability of repulsive couplings in another layer of a multiplex network.

Recent literature has indicated a strong connection between the occurrence of chimera and various responses of neurons in brain networks. For example, chimera state has been related to uni-hemispheric sleep in mammals where half of the brain remains asleep while other half remains active~\cite{uni_sleep}. This hybrid
dynamics is akin to the coexistence of coherent and incoherent spatio-temporal patterns of chimera state.
Moreover, various brain diseases have been linked to chimera states~\cite{brain_disease}. Andrzejak et.al. has shown that spatio-temporal correlation profiles, obtained from EEG readings of ecliptic seizures, bore striking similarities with hybrid patterns of chimera state~\cite{EEG_chim}. Here we study another aspect of neuronal dynamics, which is inhibition. Inhibition plays a crucial role in various cognitive abilities of brain~\cite{inhi}. Furthermore, inhibition has also been considered in ecological networks to interpret complex predator-prey interactions  among various species~\cite{inhi_ecol}. We introduce inhibition (repulsive couplings) in one layer~\cite{inhi_neg} and investigate its impact on the emergence of chimera state in another layer.

\section{Theoretical framework.}
We consider a multiplex ring network where topology of individual layer is represented by a 1D lattice with periodic boundary conditions (Fig.~\ref{multiplex}). We use Kuramoto oscillators with diffusive coupling to showcase our findings. Architecture of the multiplex network can be {represented as,}
\begin{equation}
   A=
      \begin{pmatrix} A^{(1)} & I \\ I & A^{(2)} \end{pmatrix},
\label{mul_mat}
\end{equation}
for a bi-layer multiplex network where $A^{(1)}(A^{(2)})$ represents adjacency matrix of the first (second) layer consisting of $1$ or $0$ entries depicting connected or disconnected pairs of nodes respectively. {Dynamical} evolution of the system can be expressed by a state vector $\Theta$ consisting of components $\theta_i$, $i=1,..., 2N$ representing phase of  $i^{th}$ oscillator. Dynamical equation of the network state integrating the network topology can be written as follows~\cite{chim_def}
\begin{equation}
\dot{\theta_i}= \omega_i + \lambda \sum_{j=1}^{N} A_{ij}( sin(\theta_j - \theta_i + \alpha) )
\label{eq.evol}
\end{equation}
where $\omega_i$ depicts natural frequency of the $i^{th}$ oscillator and $\lambda$ represents the  strength of the diffusive coupling. $\alpha$ is a constant phase lag parameter required for emergence of chimera state in a ring of identical phase oscillators~\cite{chim_def}. We have considered a constant natural frequency $\omega=\omega_i$ $\forall i$ for all oscillators to maintain identicality of the coupled units, a pre-requisite of definition of chimera states.

We study behavior of chimera state in the first layer ($A^{(1)}$) in the presence of repulsive couplings among same pair of nodes in the second layer ($A^{(1)}$). In our multiplex framework, the first layer is attractively coupled (i.e. with all positive entries in the corresponding adjacency matrix $A^{(1)}$) and second layer is  repulsively coupled (i.e. with all negative entries in the adjacency matrix $A^{(2)}$). Henceforth, to avoid confusion, we will refer the first layer $A^{(1)}$ as positive layer and the second layer $A^{(2)}$ as negative layer. As par with the definition of the multiplex network (Eq.~\ref{mul_mat}), we consider simple positive one to one coupling (identity matrix) for inter-layer interactions throughout the paper. Furthermore, we introduce inhibitory nodes in the second layer ($A^{(2)}$) by selecting a node (a row in $A^{(2)}$) with an inhibition probability ($p_{in}$). This leads to change  in all the coupling associated to that node to inhibitory coupling (changing the signs of each $1$ entry to $-1$ in the corresponding row of $A^{(2)}$). {For an arbitrary chosen
inhibitory node $i$, }
\begin{figure}[t]
 \centerline{\includegraphics[width=3.0in, height=2.0in]{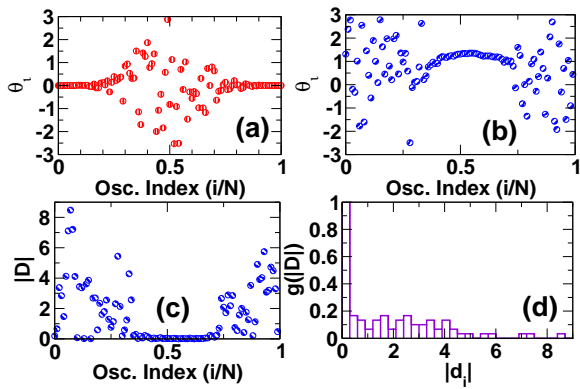}}
 \caption{ {(Color online)} (a) a realization of the initial condition which is taken from a uniform random distribution multiplied by a Gaussian profile . The Gaussian is of the form $\theta_i(t=0)=exp[-30(\frac{i}{N}-\frac{1}{2})]$. Same initial condition is considered for both the layers of multiplex 1D-1D network, (b) Snapshot of spatial phase profile of kuramoto oscillators of  the first layer of the multiplex network consisting of two attractively coupled layers, (c) Laplacian distance measure $|\bar{D}|$ of the spatial phase profile (d) normalized probability distribution function $g(| \bar{D}|)$ of the Laplacian distance measure $|\bar{D}|$. Parameters: Network size $N=N^1=N^2=100$, node degree $\langle k^1 \rangle = \langle k^2 \rangle =64$, coupling strength $\lambda=1.29$, natural frequency $\omega=0.5$ and lag parameter $\alpha=1.45$.}
 \label{numerics}
 \end{figure}
\begin{equation}\nonumber
\label{eq.1}
A^{(2)}_{i,j}=
   \begin{cases}
     -1  & \quad \text{if $i \sim j$}\\
     0 & \quad \text{otherwise}
   \end{cases}
\end{equation}
We study occurrence of the chimera state in the first layer  ($A^{(1)}$) as the introduction of inhibition breaks the $S^1$ symmetry of the second layer  ($A^{(2)}$) except for $p_{in}=1$ case where all couplings are negative. Due to this symmetry breaking in negative layer, chimera state cannot be defined in the classical sense. We have characterized the chimera state by studying the coexistence of spatial coherence and incoherence. The dynamical state of the network of phase oscillators can be mapped using the global order parameter. However, to capture varying local dynamics for chimeras, we adopt a correlation measure apt of identifying chimera states. We use a normalized probability distribution function $g(|D|)$ of the Laplacian distance measure $|D|$ and the correlation measure~\cite{chim.class}
\begin{equation}
g_0= \int_{0}^{\delta} g (|D|)  d(|D|)
\label{eq.dmsr}
\end{equation}
where $|D| =\nabla^2_i \theta= \{d_i(t): d_{i}(t)=| (\theta_{i+1}(t)-\theta_{i}(t))-(\theta_{i}(t)-\theta_{i-1}(t))|\}$ depicts presence of  a local curvature (signifying incoherence) in an otherwise smooth spatial profile (signifying spatial coherence (Fig.~\ref{numerics}) . $\delta$ is a small threshold value that sets a clear boundary between the coherent and incoherent states. Effectively, $g_0$ represents effective size of the coherent region in the spatial profile of $\theta$ at a particular time. The value of $g_0$ takes $0$ for a complete incoherent state and $1$ for the complete coherent state. A value between $0<g_0<1$ theoretically signifies existence of coherence-incoherence i.e. the chimera state~\cite{sapta_nonidentical}. Though, the correlation measure $g_0$ identifies the chimera state,
certain chimera states, for instance, breathing~\cite{breathing_c} or traveling~\cite{travelling_c}
chimera, are known to depict regular repetitive patterns in the course of time evolution.
A snapshot illustrates the chimera profile for only a fixed time point. To overcome this
constraint, we have considered the average of the correlation measure $g_0$ over 1000 consecutive time steps after an initial transient. Further, we have used a uniformly distributed random number with an Gaussian envelop to satisfy special initial condition requirement of the chimera state for Kuramoto Oscillators as depicted in Fig.~\ref{numerics}(a).

\section{Results.}
{We present results for dynamical evolution of coupled Kuramoto oscillators on multiplex networks with both layers represented by 1D lattices. In the first layer, all the nodes are attractively coupled, whereas in {the} second layer the nodes are connected via either attractive or repulsive coupling with {the probability} $p_{in}$ deciding the population of inhibitory nodes.}
\begin{figure}[b]
 \centerline{\includegraphics[width=3.5in, height=1.5in]{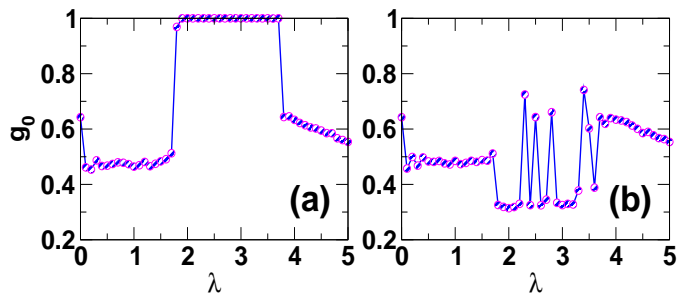}}
 \caption{ {(Color online)} Normalized probability distribution function $g(| \bar{D}|)$ for the Laplacian distance measure $|\bar{D}|$  of  the first layer of the 1D-1D multiplex network consisting of (a) two positive layers,  and (b) positive-negative layers. Parameters: Network size $N=N^1=N^2=100$, node degree $\langle k^1 \rangle = \langle k^2 \rangle = 64$,  natural frequency $\omega=0.5$ and lag parameter $\alpha=1.45$.}
 \label{npb_pp_ps}
 \end{figure}

\subsection{Multiplexing with a repulsively coupled layer.}
{First, we consider a multiplex network where in one layer all nodes are positively (attractively) coupled and in another layer all nodes are negatively (repulsively) coupled}. We particularly compare dynamical state of the positive layer for the {following} two cases; (I) a multiplex network consisting of two layers with attractive couplings in both the layers, and (II) a multiplex network with one attractive and one repulsive layer. For the case (II), all the entries {in} $A^{(2)}$ are negative corresponding to $p_{in}=1$. We study changes in {the} oscillator dynamics in {the} positive layer when it is multiplexed with another positive layer (case (I)), with {that of} negative layer (case (II)). Note that, due to {the} symmetric coupling environment for both the layers in the multiplex network, we can define chimera state {for} both the layers separately in this particular combination (i.e. {for} $p_{in}=1$). Fig.~\ref{npb_pp_ps} {plots the} correlation measure {indicating} the range for appearance of chimera state as a function of coupling strength. We find that {for} lower coupling {values}, {for both the cases}, positive layer shows chimera state as depicted by $0<g_0<1$ values of the correlation measure (Fig.~\ref{npb_pp_ps}). {It should be noted that} a high value of {the correlation measure} ($g_0\approx0.64$) at coupling strength $\lambda=0$ arises from the fact that without the coupling, the oscillators evolve with their constant natural frequency resulting in {the} same spatial profile {considered for} the special initial condition.

\begin{figure}[t]
 \centerline{\includegraphics[width=\columnwidth]{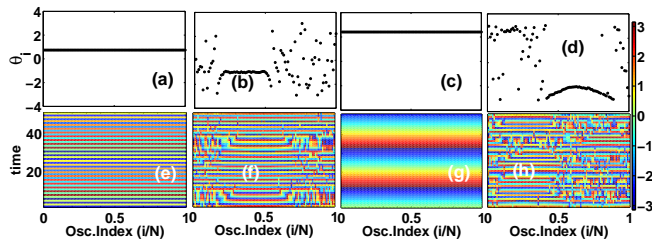}}
 \caption{ {(Color online)}  Snapshots and spatio-temporal patterns depicting emergence of chimera in positive layer upon multiplexing with a negative layer. The 1D-1D multiplex network consists of (a,c,e,g) two positively coupled layer, (b,d,f,h) positively and negatively couples layers. (a,b,c,d) presents a snapshot of the spatio-temporal patterns presented in (e,f,g,h) respectively. The figures represents emergence of chimera in case of multiplexing with negative layer at coupling strength $\lambda= 1.86$ (a,b) and at $\lambda= 3.7$ (c,d). Other parameters are same as Fig.~\ref{npb_pp_ps}}
 \label{snapshots}
 \end{figure}

Interestingly, we observe a contrasting behavior in the middle coupling range ($2 \lesssim\lambda \lesssim 4$). For the case (I), the oscillators demonstrate a transition to the synchronized state represented by $g_0=1$ whereas the case (II) demonstrates a intermediate correlation value ($0.3 \lesssim g_0 \lesssim 0.7$) representing the chimera state. Fig.~\ref{snapshots} (a),(e) {exhibit} a completely synchronized state in the positive layer when multiplexed with another positive layer (case (I)). Fig.~\ref{snapshots}(b),(f) show a chimera state in {the} positive layer upon its multiplexing with a negative layer (case (II)). Furthermore, Fig.~\ref{snapshots}(c),(d) provide another {illustration of this} destruction of synchrony and an enhancement {in} {the} chimera state for different coupling strength. Fig.~\ref{snapshots}(e)-(h) represent the sptio-temporal patterns of the chimera state which indicates that the emerged chimera state is stable with time. Moreover, the completely synchronized state for
the case I demonstrates a periodic temporal evolution~\cite{suppl}. Replacing one positively coupled layer with a negatively coupled layer makes the stable periodic evolution unstable leading to an hybrid spatial chimera pattern~\cite{suppl}. This observation highlights the importance of {repulsively coupled layer} which causes {an occurrence {of} chimera state in a attractive coupled layer due to {the} multiplexing.}

\subsection{Controlling chimera in one layer by changing inhibitory coupling in another layer.}
{Next,} we investigate the impact of inhibitory nodes in one layer on the emergence of chimera state in another layer.
 \begin{figure}[b]
 \centerline{\includegraphics[width=3.5in, height=1.5in]{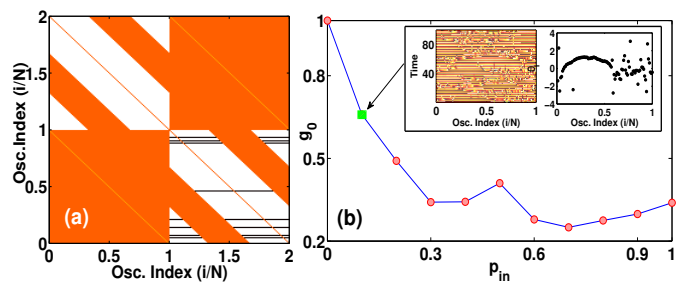}}
 \caption{{(Color online)} (a) Plot of the adjacency matrix of 1D-1D multiplex network consisting of one positive and one inhibitory layer with inhibition probability $p_{in}=0.1$. The black lines in the down-right block represents the rows corresponding to inhibitory nodes. (b) Normalized probability distribution function $g(| \bar{D}|)$ of the Laplacian distance measure $|\bar{D}|$ of  the positive layer as a function of inhibition probability $p_{in}$. Inset: Snapshot and spatio-temporal profile of the first layer of a multiplex network consisting of one positive layer and one inhibitory layer. Parameters: Coupling strength $\lambda=3.57$, Inhibition probability $p_{in}=0.1$, Other parameters are same as Fig.~\ref{npb_pp_ps}.}
 \label{pin_0.1}
 \end{figure}
{Again, we consider a bi-plex} network where first layer ($A^{(1)}$) {has} all positive couplings (thus termed as positive layer). The {inhibitory nodes are introduced} in the second layer ($A^{(2)}$) with inhibition probability $p_{in}$. Fig~\ref{multiplex} (c) depicts a schematic diagram of such arrangement in multiplex network and Fig.~\ref{pin_0.1} (a) shows plot of a multiplex adjacency matrix where inhibitory nodes (represented by black lines in right-down block) are introduced with an inhibition probability $p_{in}=0.1$. In this setup, a particular pair of nodes in $A^{(2)}$ may interact via {either positive or negative couplings decided by $p_{in}$}. We find that an introduction of even a small number of the inhibitory nodes is sufficient to destroy synchronized regime in $A^{(1)}$ {and causes an enhancement in} the range of couplings strength for which chimera is appeared in the positive layer. Fig.~\ref{pin_0.1} depicts variation of the correlation measure with respect to the inhibition probability $p_{in}$ for a {sufficient large coupling strength ($\lambda=3.57$). Two extreme situations}, i.e. $p_{in}=0$ and $p_{in}=1$ correspond to a completely synchronized state ($g_0=1$) and a chimera state ($g_0 \approx 0.34$) respectively, {which is not surprising} as discussed in the previous section. However, {the interesting fact is that} even a small amount of inhibition {in the repulsive layer} (say $p_{in}=0.1$) {is sufficient to yield a non-zero correlation value} ($g_0 \approx 0.65$) {for the attractive layer} depicting chimera state. Inset figures of Fig.~\ref{pin_0.1} {manifest this phenomena of appearance of the chimera state} at low inhibition probability.

\section{Conclusion.}
To surmise, we have explored {an} impact of {inhibitory (repulsive) coupling} in {one layer} layer on dynamical behavior of {another} layer in a multiplex network. We have systematically studied {impact} of {inhibition} by inserting inhibitory nodes with {a} probability $p_{in}$ {which is varied from {zero to one. The
case} $p_{in}$ being zero corresponds to {a multiplex network consisting} two identical layers {with each layer} having attractively coupled nodes. The another extreme case $p_{in}$ {being one} corresponds to {a multiplex network consisting of two} layers {which are structurally identical} but vary in the nature of coupling, i.e., one layer has all attractively {coupled nodes whereas} another {layer} has all repulsively {coupled nodes}. Due} to the {multiplexing with a} layer having all inhibitory nodes, synchrony among phase oscillators with all attractively couplings are destroyed. Furthermore, we {report} that the range of parameters for which chimera is demonstrated {in one layer} can be controlled by {changing} probability of inhibitory nodes in another layer. {Importantly}, we found that a very small number of inhibitory {nodes} can bring an enhancement in the {appearance} of chimera state {destroying the synchronized} state. The results presented here can be helpful to gain better understanding of the workings of brain {under different conditions} and {can help in diagnosis of several brain {related} diseases}.
Furthermore, {we have focused here on} {behavioral changes in one layer} due to
{inhibitory nodes in another layer in a multiplex framework}.
This investigation can be further extended by focusing the impact of
particular positions of inhibitory nodes on the occurrence of different spatial clusters appeared in the enhanced chimera state. Furthermore, other network topologies can be considered for the second
layer to have a better representation of real-world complex systems. These results promote
importance of the multiplex framework to model those real-world complex systems which posses
more than one type of interactions {among their} constituents.

\section{Acknowledgment}
SJ is grateful to Department of Science and Technology (DST), Government of India grant EMR/2016/001921 for financial support. SG acknowledges DST, Government of India, for the INSPIRE fellowship (IF150149) as well as the Complex Systems Lab members for timely help and useful discussions. SJ acknowledge discussions with Hildegard Meyer-Ortmanns on negative couplings during a visit to MPIPKS, Dresden. BP thanks complex systems lab @ IITI where the work was done.

\end{document}